\documentclass[prx,twocolumn, aps,letterpaper, preprintnumbers,superscriptaddress ]{revtex4-2}
\usepackage{hyperref}
\usepackage{verbatim}
\usepackage{amsmath}
\usepackage{latexsym}
\usepackage{revsymb}
\usepackage{yfonts}
\usepackage{ifthen}

\usepackage{natbib}
\usepackage{amsfonts}
\usepackage{amsmath}
\usepackage{amssymb}
\usepackage{amsthm}
\usepackage{graphicx}
\usepackage{bm}
\usepackage{bbm}
\usepackage{epsfig,color,amssymb}
\usepackage{amsfonts}
\usepackage{amscd}
\usepackage{amsmath}
\usepackage{multirow}
\usepackage{chemarrow}
\usepackage{dcolumn}
\usepackage{bm}
\usepackage{graphicx}
\usepackage{enumerate}
\usepackage{epsfig}
\usepackage{subfigure}
\usepackage{xcolor}

\usepackage{ulem}
\usepackage{braket}
\usepackage{comment}
\usepackage{enumitem}
\usepackage{amsthm}
\usepackage{ragged2e}
\justifying\let\raggedright\justifying

\renewcommand{\emph}[1]{\textit{#1}}

\def\bra#1{\mathinner{\langle{#1}|}}
\def\ket#1{\mathinner{|{#1}\rangle}}

\def\Bra#1{\left\langle{#1}\right|}
\def\Ket#1{\left|{#1}\right\rangle}
\def\fid#1#2{\left\langle{#1}|{#2}\right\rangle}

\usepackage{amsfonts}
\usepackage{amsmath}
\usepackage{amsthm}
\usepackage{amscd}
\usepackage{amssymb}
\usepackage{amsxtra}
\usepackage{gensymb}
\usepackage{epstopdf}
\usepackage{color}
\usepackage{makecell}
\usepackage{multirow}
\usepackage{stackengine}
\usepackage{threeparttable}
\usepackage{upgreek}

\newcommand\xrowht[2][0]{\addstackgap[.5\dimexpr#2\relax]{\vphantom{#1}}}
\hypersetup{colorlinks=true, linkcolor=blue, citecolor=red, urlcolor=blue}
\begin{document}
	\title{Experimental measurement-device-independent type quantum key distribution with flawed and correlated sources}

	\author{Jie Gu}\thanks{These authors contributed equally to this work}
	\author{Xiao-Yu Cao}\thanks{These authors contributed equally to this work}
	\affiliation{National Laboratory of Solid State Microstructures and School of Physics, Collaborative Innovation Center of Advanced Microstructures, Nanjing University, Nanjing 210093, China}
	\author{Yao Fu}
	\affiliation{Beijing National Laboratory for Condensed Matter Physics and Institute of Physics, Chinese Academy of Sciences, Beijing 100190, China}
	\affiliation{MatricTime Digital Technology Co. Ltd., Nanjing 211899, China}
	\author{Zong-Wu He}
	\affiliation{National Laboratory of Solid State Microstructures and School of Physics, Collaborative Innovation Center of Advanced Microstructures, Nanjing University, Nanjing 210093, China}
	\affiliation{State Key Laboratory of Particle Detection and Electronics, University of Science and Technology of China, Hefei 230026, China}
	\author{Ze-Jie Yin}
	\affiliation{State Key Laboratory of Particle Detection and Electronics, University of Science and Technology of China, Hefei 230026, China}
	\author{Hua-Lei Yin}\email{hlyin@nju.edu.cn}
	\affiliation{National Laboratory of Solid State Microstructures and School of Physics, Collaborative Innovation Center of Advanced Microstructures, Nanjing University, Nanjing 210093, China}
	\author{Zeng-Bing Chen}\email{zbchen@nju.edu.cn}
	\affiliation{National Laboratory of Solid State Microstructures and School of Physics, Collaborative Innovation Center of Advanced Microstructures, Nanjing University, Nanjing 210093, China}
	\affiliation{MatricTime Digital Technology Co. Ltd., Nanjing 211899, China}

	\begin{abstract}	
		The security of quantum key distribution (QKD) is severely threatened by discrepancies between realistic devices  and theoretical assumptions. Recently, a significant framework called the reference technique was proposed to provide security against arbitrary source flaws under current technology such as state preparation flaws, side channels caused by mode dependencies, the Trojan horse atttacks and pulse correlations. Here, we adopt the reference technique to prove security of an efficient four-phase measurement-device-independent QKD using laser pulses against potential source imperfections. We present a characterization of source flaws and connect them to experiments, together with a finite-key analysis against coherent attacks. In addition, we demonstrate the feasibility of our protocol through a proof-of-principle experimental implementation and achieve a secure key rate of 253 bps with a 20 dB channel loss. Compared with previous QKD protocols with imperfect devices, our study considerably improves both the secure key rate and the transmission distance, and shows application potential in the practical deployment of secure QKD with device imperfections.\\
		
	\end{abstract}

	\maketitle

\section{INTRODUCTION}
Quantum key distribution (QKD) enables two distant participants, Alice and Bob, to share secure key bits~\cite{bennett1984proceedings,Ekert1991Quantum} for the encryption and decryption of secret communication. Together with the one-time pad algorithm, QKD offers theoretical security for information exchanges based on the laws of quantum mechanics~\cite{Xu2020Secureqkd,pirandola2020advances}. However, for the practical operation of QKD systems, a serious security loophole still exists, which occurs due to the discrepancy between theoretical security assumptions and practical devices. To be precise, a security proof of QKD is established with assumptions on the system devices, which cannot be satisfied for realistic devices because of inherent imperfections and the disturbance of eavesdroppers  ~\cite{lydersen2010hacking,tang2013source,xu2010experimental,Huang2019Laser-seeding}. This deviation results in more information leakage to eavesdroppers, which cannot be noticed by users~\cite{Xu2020Secureqkd}.

To narrow the discrepancy  and further strengthen the security against device flaws, some crucial protocols have been presented, one of which is device-independent QKD, which guarantees the unconditional security of QKD by measuring the violation of Bell's inequality~\cite{PhysRevLett.98.230501}. Although a number of experimental demonstrations of the loophole-free Bell test have been conducted~\cite{Hensen2015,Shalm2015Loophole-free,PhysRevLett.115.250401}, experimental implementations of device-independent QKD still suffer from short transmission distances, because they require nearly perfect single-photon detection. To date, several device-independent QKD experiments have been performed, which show that device-independent QKD is still far from being implemented in long-distance transmission~\cite{Nadlinger2022,Zhang2022,PhysRevLett.129.050502}, as listed in Table~\ref{tab01}. 
\begin{table}[t]
	\centering
	\caption{Comparison of the experimental performance for different methods for QKD against source and detection flaws.}\label{tab01}
	\setlength{\tabcolsep}{0.001cm}
	\begin{threeparttable}
		\center\begin{tabular}{ccc}\hline \xrowht[()]{20pt}
			& \makecell[c]{Performance\tnote{2}} & Key rate (bps)\\
			\hline \xrowht[()]{15pt}
			This study &10 dB attenuation & $9.10\times 10^{4} $\\
			& 20 dB attenuation & 253  \\
			\xrowht[()]{15pt}
			Side-channel-secure\tnote{1}~\cite{zhang2022experimental}& \makecell[c]{50 km fiber spool} & 173\\
			\xrowht[()]{15pt}
			Device-independent~\cite{Nadlinger2022} &\makecell[c]{ 2 m spatial distance} & $\sim$ 7.43 \\
			\xrowht[()]{15pt}
			Device-independent~\cite{Zhang2022} & \makecell[c]{400 m spatial distance}&$8.5\times 10^{-4} $ \\
			\xrowht[()]{15pt}
			Device-independent~\cite{PhysRevLett.129.050502} & 220 m fiber spool & 2.60  \\
			\hline
		\end{tabular}
		\begin{tablenotes}
			\footnotesize
			\item[1] Ref.~\cite{zhang2022experimental} is a side-channel-secure QKD experiment (assuming perfect vacuum and without considering pulse correlations).\\
			\item[2] The performance of QKD experiments can be demonstrated by realistic transmission or attenuation simulation. The attenuation of 50 km fiber spool is about 10 dB.
		\end{tablenotes}
	\end{threeparttable}
\end{table}Furthermore, a recent study revealed that Bell nonlocality is insufficient to prove the security of standard device-independent QKD protocols~\cite{PhysRevLett.127.050503} in the large-noise regime.

Another significant approach  towards the security of practical QKD is measurement-device-independent (MDI) QKD~\cite{Lo2012MDI,Braunstein:2012:Side-Channel-Free,PhysRevA.87.012320,Curty2014,PhysRevA.89.052333,PhysRevA.91.032318}, as well as its single-photon interference version, twin-field QKD~\cite{Lucamarini2018}, which closes all attacks on measurement devices. Both of them are practical~\cite{Rubenok2013real-worldtp,tang2014measurement,Wang2015Phase,zhou2016making,PhysRevLett.112.190503,Comandar2016,zheng2021heterogeneously,xie2021breaking}, and break the transmission record using current technologies~\cite{Yin2016MDI404,Fang2020impltf,Pittaluga2021,Chen2021,Wang2022}. Espeically, twin-field QKD protocols can break the Pirandola-Laurenza-Ottaviani-Banchi (PLOB) bound, which is the repeaterless secret key capacity bound~\cite{PhysRevLett.102.050503,Pirandola2017}. However, source flaws present a loophole  towards securing QKD with imperfect devices. In recent years, several studies have been conducted to improve the security with source flaws, such as the loss-tolerant protocol~\cite{PhysRevA.90.052314,Xu2015Experimental,Tang2016Exp,pereira2019quantum} and other attempts~\cite{Braunstein:2012:Side-Channel-Free,Yin2013Mea,PhysRevA.104.022423,PhysRevApplied.12.054034,Li2014Quantum,ding2022measurement,huang2022characterisation}.
Recently, an important theoretical study was published to propose a security analysis method for practical QKD against source flaws, which is called the reference technique (RT)~\cite{PhysRevApplied.15.034072,Pereiraeaaz4487,PhysRevA.104.062611}. To be precise, the RT is a parameter estimation technique which can help simplify the estimation of parameters needed for security proof through several reference states. Using the RT method, parameters characterizing typical source flaws, such as state preparation flaws (SPFs)~\cite{jiang2021robust}, side channels caused by mode dependencies, Trojan horse attacks (THAs), and classical pulse correlations~\cite{PhysRevA.104.062611}, can be estimated to guarantee the security of the actual QKD protocol.

Here, we adopt the RT method to prove the practical security of a four-phase (FP) MDIQKD protocol~\cite{PhysRevA.85.042307} using coherent light sources against all possible source flaws, including pulse correlations. We apply the setup of phase encoding schemes~\cite{PhysRevA.85.042307} and the single photon interference introduced in twin-field QKD~\cite{Lucamarini2018} contributes to improving the key rate of our scheme. In addition, we perform a proof-of-principle experimental implementation of our protocol to demonstrate its feasibility. Specifically, we fully characterize imperfect sources using the state representation in Ref.~\cite{Pereiraeaaz4487} (the reference state representation) and introduce bounded fidelity to offer a secure key rate by using the RT method. Although bounded fidelity was first introduced in the Gottesman-Lo-Lütkenhaus-Preskill (GLLP) analysis~\cite{lo2007security,Koashi_2009}, the GLLP analysis can be included under the RT method as a special case~\cite{Pereiraeaaz4487}. We simulated the secure key rate of our protocol and compared it with protocols in Refs.~\cite{PhysRevApplied.15.034072} and ~\cite{PhysRevApplied.12.054034} which are both against source flaws. The simulation results show that both in the ideal scenario and in the scenario with source flaws, the secure key rate of our protocol shows a higher performance than the other two protocols.

We specify the parameters characterizing source flaws in the reference state representation and further relate them to data that can be measured experimentally to help generate a secure key rate in the experiment. Our experimental implementation only requires four-phase (quadrature phase-shift keying ~\cite{Liu2021Hom}) modulation of coherent states, which avoids intensity modulation correlations~\cite{zapatero2021security} and does not require vacuum state preparation. In addition, we conducted a finite-key analysis of our protocol, which is secure against coherent attacks. Considering source flaws, our experiments can achieve a key rate of 253 bps with a 20 dB channel loss, which is a considerable progress compared with previous experimental implementations of QKD protocols with imperfect source and measurement devices~\cite{zhang2022experimental,Nadlinger2022,Zhang2022,PhysRevLett.129.050502}. Our study is the first experimental implementation of a QKD protocol using the RT method~\cite{pereira2019quantum,Pereiraeaaz4487},  with practical security against currently all possible imperfections in devices. The method we used to build relations between parameters estimated through the RT method and experimental data is also compatible with other potential source imperfections. Our study proves the feasibility of a practical secure QKD using the RT method. The features of our protocol also jointly show possible practicality for the future deployment of secure QKD systems with imperfect devices .

\section{Materials and methods}
\subsection{Protocol description}
\begin{figure}
	\centering
	\includegraphics[width=8.6cm]{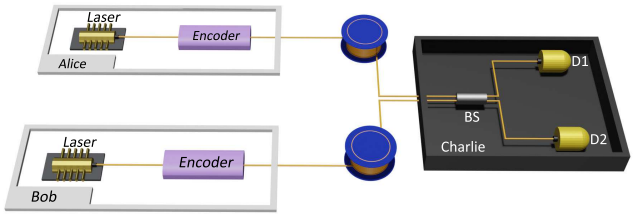}\\
	\caption{Schematic of our FP-MDIQKD protocol. At Alice's (Bob's) site, the simplified transmitter contains a laser source and an encoder. The encoder modulates the phase of the optical pulses. The modulation information of the encoder should be monitored to characterize the deviation between ideal  and realistic pulses. If the untrusted Charlie is honest, he measures the signals received using an interference measurement.}\label{fig01}
\end{figure}

A brief schematic of the FP-MDIQKD protocol is shown in Fig.~\ref{fig01}. In our QKD system, Alice and Bob send weak coherent state pulses to an untrusted node, Charlie, who performs the interference measurement. For simplicity, we first describe our protocol under the assumption that Alice and Bob are able to prepare perfect optical pulses. Source flaws are thoroughly discussed afterwards.
The protocol comprises the following steps:

(i) \emph{Preparation.} In each turn, Alice randomly chooses the $\rm X$ and $\rm Y$ bases with probabilities $p_{x}$ and $p_{y}=1-p_{x}$, respectively. If selecting the $\rm X$ basis, she picks a "key" bit $k_{x}^{a}\in\{0,1\}$ and generates a weak coherent state pulse $\ket{e^{ik_{x}^{a}\pi}\alpha}$, where $|\alpha|^2$ is the intensity of the optical pulse. When choosing the $\rm Y$ basis, she prepares a weak coherent state pulse $\ket{e^{i(k_{y}^{a}+\frac{1}{2})\pi}\alpha}$ according to random bit $k_{y}^{a}\in\{0,1\}$. Bob does the same.

(ii) \emph{Measurement.} Alice and Bob send their optical pulses to an untrusted middle node Charlie through insecure quantum channels. Charlie is expected to perform interference measurements and record the detector that clicks.

Note that Charlie is assumed to apply a 50:50 beam splitter to the incoming two optical pulses, followed by two threshold single-photon detectors, D1 and D2. We assume that D1 (D2) will click when the phase difference between the two incoming optical pulses is zero ($\pi$). We assume that Charlie records the situation when one and only one detector clicks as an effective measurement, regardless of the actual clicking results, because Charlie is an untrusted node and his betrayal can be included in the security analysis.

(iii) \emph{Sifting.} Alice and Bob repeat Steps (i)–(ii) many times. Then, Charlie announces the effective measurements and  detection results (whether D1 or D2 clicks). Alice and Bob maintain their corresponding keys and basis choices based on Charlie's announcement. If Charlie claims that D2 clicks, Bob will flip his key bit. Alice and Bob then disclose their basis choices for effective measurements through authenticated classical channels, and further classify their key bits with basis information.

(iv) \emph{Parameter estimation.} For the retained key bits in the $\rm X$ and $\rm Y$ bases, Alice and Bob obtain the gain. Then, they disclose all key bits under the $\rm Y$ basis to calculate the bit error rate, which is used to set a bound on the phase error rate.

(v) \emph{Key distillation.} Alice and Bob perform error correction and privacy amplification on the remaining sifted keys under the $\rm X$ basis to distill the final secure key bits.

Evidently, the use of weak coherent state pulses naturally makes it easier for us to characterize the SPFs because theoretically any deviation in intensity and phase modulation between ideal pulses and realistic pulses can be directly shown in the expression of laser pulses. No assumption on dimensions of Hilbert spaces is required. Our protocol further inherits the advantages that the QKD system can be implemented with fixed intensity because a recent study reveals that intensity modulation introduces extra pulse correlations of intensity~\cite{zapatero2021security}.
	
\subsection{Key rate of FP-MDIQKD}
As depicted in Fig.~\ref{fig01}, the setting of two bases can introduce bounded fidelity with imbalance $\Delta$ of a quantum coin~\cite{PhysRevA.85.042307,PhysRevA.86.059903,lo2007security}. For simplicity, we only present the main result for key rate estimation because the idea of a quantum coin has been discussed in previous studies~\cite{lo2007security,Koashi_2009} (see Supplementary Material). Because only key bits under the $\rm X$ basis are used to form secure key bits, the final key rate of our FP-MDIQKD protocol is given by
\begin{equation}\label{key}
	R = Q^{x}[1-fH(E_{\rm b}^{x})-H(E_{\rm p})]
\end{equation}
where $Q^{x}$ denotes the gain under the X-basis. $E_{\rm p}$ is the phase error rate under the X-basis. $E_{\rm b}^{x}$ represents the quantum bit error rate under the X basis. $H(x)=-x\log_{2}x-(1-x)\log_2(1-x)$ is the binary Shannon entropy function. According to the GLLP analysis and the RT method~\cite{lo2007security,Koashi_2009,PhysRevA.85.042307,PhysRevA.86.059903,Pereiraeaaz4487}, the phase error rate $E_{\rm p}$ can be bounded by
\begin{equation}
	1-2\Delta\leq\sqrt{E_{\rm b}^{y}E_{\rm p}}+\sqrt{(1-E_{\rm b}^{y})(1-E_{\rm p})},
\end{equation}
where $E_{\rm b}^{y}$ is the quantum bit error rate in the $\rm Y$ basis and $\Delta$ quantifies the imbalance of Alice's and Bob's quantum coins based on their basis choices. $\Delta$ can be bounded by the fidelity of the two basis-dependent states under the X and Y bases, which is denoted as $|\fid{\Psi_{\rm Y}}{\Psi_{\rm X}}|$. In the symmetric scenario in which Alice and Bob share basis-dependent states, the simplified relation between $\Delta$ and fidelity is~\cite{PhysRevA.86.059903} $1-2Q\Delta={\rm Max}_{\theta^{\prime}, \delta_{\rm X}, \delta_{\rm Y}}{\rm Re}\left(e^{i\theta^\prime}\fid{\Psi_{{\rm Y}, \delta_{\rm Y}}}{\Psi_{{\rm X}, \delta_{\rm X}}}\right)|\fid{\Psi_{\rm Y}}{\Psi_{\rm X}}|$, where $Q$ is the total gain. Here, $\theta^\prime, \delta_{\rm X}$, and $\delta_{\rm Y}$ are free variables ranging from 0 to $2\pi$ and $\Ket{\Psi_{{\rm V}, \delta_{\rm V}}}=(\Ket{0_{\rm V}}\Ket{\Psi_{0\rm V}}+\Ket{1_{\rm V}}\Ket{\Psi_{1\rm V}})/\sqrt{2}$. $\Ket{\Psi_{j\rm V}}$ is the state that Alice or Bob actually prepares in the implementation under the V basis (V = X or Y) with the bit value $j$. For the experimental implementation, the calculation of fidelity with imperfect sources can be expressed as
\begin{equation}\label{practicalfinal}
	\begin{aligned}
		\fid{\Psi_{\rm X}}{\Psi_{\rm Y}}=&\frac{1}{4}(1-\epsilon)e^{-\mu}{\rm cos}^2\theta\lbrack(1-i)\fid{\alpha^\prime}{ie^{i\delta}\alpha^\prime}\\
		&+(1-i)\fid{-e^{i\delta}\alpha^{\prime}}{-ie^{i\delta}\alpha^{\prime}}\\
		&+(1+i)\fid{\alpha^{\prime}}{-ie^{i\delta}\alpha^{\prime}}\\
		&+(1+i)\fid{-e^{i\delta}\alpha^{\prime}}{ie^{i\delta}\alpha^{\prime}})\rbrack,
	\end{aligned}
\end{equation}
where we use $\epsilon$, $\mu$, and $\theta$ to characterize pulse correlations, THAs, and side channels caused by mode dependencies, respectively. Here, we consider special cases of THAs and side channels as discussed in the reference state representation~\cite{pereira2019quantum}. For the THAs, $\mu$ is the intensity of the back-reflected light sent by the eavesdroppers. For the side channels, $\theta$ is the parameter used to bound the side channels in the polarization space of our protocol in Eq.~\eqref{sidechannel}. Note that other potential side channels can also be characterized by additional parameters. For SPFs, we consider only the worst scenario with the maximum deviation in intensity and phase modulation without pulse correlations. $|\alpha^\prime|^2$ represents the realistic intensity modulation of coherent pulses. The maximum deviation ratio can be denoted by the optical power fluctuation $\xi$ and $\xi=\left||\alpha^\prime|^2-|\alpha|^2\right|/|\alpha|^2$. $\delta$ represents the deviation from the ideal phase modulation. Aside from the THAs, the parameters of the side channels and SPFs discussed here can be directly measured in the experiment, which can be seen in our experimental implementation. By contrast, pulse correlations are usually witnessed during experiments in the form of pattern effects~\cite{Yoshino2018}, which means $\epsilon$ cannot be directly derived through experimental measurements. Since our protocol was one FP-MDIQKD with fixed intensity, the pattern effect occurs in the phase modulation, which is denoted as $\psi$ and $\psi$ represents the maximum deviation of the phase modulation caused only by pulse correlations. Here, we derive  a formula that relates $\epsilon$ to $\psi$, which can be obtained as
\begin{equation}
	\epsilon=1-e^{|\alpha|^2(2{\rm cos}\psi-2)}.
\end{equation}
We remark that $\epsilon$ is the parameter generalizing arbitrarily long-range pulse correlations. In our experiments, we only measure pattern effects happening between nearest-neighbor pulses. However, our method can also be easily applied to measure long-range pattern effects when the length of pulse correlations is more than one, as shown in Supplementary Material.

For experimental implementation, we provide a finite-key analysis of our protocol based on the method introduced in Ref.~\cite{Lorenzo2021} and the analysis  is secure against coherent attacks. The formula for the number of secure key bits $l$ can be expressed as
\begin{equation}\label{finite}
	\begin{aligned}
		l=n_{x}\Bigg[1 & -H(\overline{E}_{\rm p})-{\rm leak}_{\rm EC}                                                                       \\
		               & -\frac{1}{n_x}{\rm log}_2\frac{2}{\epsilon_{\rm EC}}-\frac{1}{n_x}{\rm log}_2\frac{1}{4\epsilon^2_{\rm PA}}\Bigg],
	\end{aligned}
\end{equation}where $n_x$ denotes the number of raw bits under the X basis, which is used to generate the secret key, and ${\rm leak}_{\rm EC}=fH(E_{\rm b}^{x})$ represents the percentage of bits consumed during the error correction. $\epsilon_{\rm EC}$ and $\epsilon_{\rm PA}$ represent the failure probabilities for the error correction and privacy amplification, respectively. $\bar{\epsilon}$ is the other error probability that measures the accuracy of estimating smooth min-entropy. In addition, statistical fluctuations should be applied to determine the upper bound of the observed phase-error rate $\overline{E}_{\rm p}$. Detailed formulas can be found in the Supplementary Material.
	
\subsection{Characterization of source flaws}
We characterized the expression for basis-dependent states in our protocol. First, an ideal scenario is presented. Then, we discuss how to jointly consider different realistic source flaws (SPFs, side channels arising from mode dependencies, THAs and {pulse} correlations) to describe the basis-dependent states using the reference state representation.

\subsubsection{The ideal scenario}
In the ideal scenario, Alice and Bob prepare perfect weakly coherent states. The symmetric scenario is another assumption under which the states Alice and Bob prepare are the same. The basis-dependent states are
\begin{equation}
	\begin{aligned}
		&\Ket{\Psi_{\rm X}}=\frac{1}{\sqrt{2}}\left( \Ket{0_{\rm X}}\Ket{\alpha}+\Ket{1_{\rm X}}\Ket{-\alpha} \right),\\
		&\Ket{\Psi_{\rm Y}}=\frac{1}{\sqrt{2}}\left( \Ket{1_{\rm Y}}\Ket{i\alpha}+\Ket{0_{\rm Y}}\Ket{-i\alpha} \right),\\
	\end{aligned}
\end{equation}
where $\{\Ket{0_{\rm X}}, \Ket{1_{\rm X}}\}$ and $\{\Ket{0_{\rm Y}}, \Ket{1_{\rm Y}}\}$ are the eigenstates of the Pauli operators $\sigma_x$ and $\sigma_y$, respectively.

\subsubsection{State preparation flaws}
When considering SPFs in practical implementations, there is an inherent advantage of using weak coherent pulses as SPFs can be characterized by imperfect intensity and phase modulations on weak coherent pulses~\cite{jiang2021robust}. The actual basis-dependent states with SPFs can be expressed as
\begin{equation}
	\begin{aligned}
		&\Ket{\Psi_{\rm X}}^{*}=\frac{1}{\sqrt{2}}\left( \ket{0_{\rm X}}\ket{\alpha_0}+\ket{1_{\rm X}}\ket{-e^{i\delta_1}\alpha_1} \right),\\
		&\Ket{\Psi_{\rm Y}}^{*}=\frac{1}{\sqrt{2}}\left( \ket{1_{\rm Y}}\ket{ie^{i\delta_2}\alpha_2}+\ket{0_{\rm Y}}\ket{-ie^{i\delta_3}\alpha_3} \right).
	\end{aligned}
\end{equation}
Note that the phase in $\ket{\alpha_0}$ can be regarded as the global phase and does not affect the bounded fidelity of the two bases. Here, we consider a situation with varied basis and bit choices, which have different deviations of intensity and phase modulations,  to derive  the general expression of the basis-dependent states. For the simulation and experimental implementations, as mentioned before, we only considered the worst scenario with maximum deviations in intensity and phase modulations without {pulse} correlations.

\subsubsection{Side channels caused by mode dependencies}
Side channels caused by mode dependencies are common imperfections in a practical QKD system, where the eavesdropper Eve can exploit dimensions other than the encoding dimension to distinguish states that carry the secret. For example, in our protocol, the bits are encoded in phase space, which is the operational space, and Eve can distinguish signals from senders using the polarization information of each state, which is one of the side channels . Here, we consider only the polarization space as the side-channel space, together with a polarization multimode scenario. We express the weak coherent states according to the basis and bit information as follows:
\begin{equation}\label{sidechannel}
	\begin{aligned}
		&\ket{\alpha_0^{\prime}}=\rm cos\theta_0\ket{\alpha_0}_{\rm H}+\rm sin\theta_0\ket{\alpha_0}_{\rm V},\\	
		&\ket{\alpha_1^{\prime}}=\rm cos\theta_1\ket{-e^{i\delta_1}\alpha_1}_{\rm H}+\rm sin\theta_1\ket{-e^{i\delta_1}\alpha_1}_{\rm V},\\
		&\ket{\alpha_2^{\prime}}=\rm cos\theta_2\ket{ie^{i\delta_2}\alpha_2}_{\rm H}+\rm sin\theta_2\ket{ie^{i\delta_2}\alpha_2}_{\rm V},\\
		&\ket{\alpha_3^{\prime}}=\rm cos\theta_3\ket{-ie^{i\delta_3}\alpha_3}_{\rm H}+\rm sin\theta_3\ket{-ie^{i\delta_3}\alpha_3}_{\rm V},\\
	\end{aligned}
\end{equation}
where H and V refer to the horizontal and vertical polarization modes, respectively. In addition, we assume that $\theta_0=\theta_1=\theta_2=\theta_3=\theta$, as we consider the worst situation when the maximum deviation of polarization $\theta$ occurs for every state. We adopt only states in the horizontal polarization mode to form the secure key rate, and $\theta$ can be directly measured through the extinction ratio of polarization. We obtain the basis-dependent states with side channels and SPFs as
\begin{equation}
	\begin{aligned}
		&\Ket{\Psi_{\rm X}}^{\prime}=\frac{1}{\sqrt{2}}\left( \ket{0_{\rm X}}\ket{\alpha_0^{\prime}}+\ket{1_{\rm X}}\ket{\alpha_1^{\prime}} \right),\\
		&\Ket{\Psi_{\rm Y}}^{\prime}=\frac{1}{\sqrt{2}}\left( \ket{1_{\rm Y}}\ket{\alpha_2^{\prime}}+\ket{0_{\rm Y}}\ket{\alpha_3^{\prime}} \right).\\	
	\end{aligned}
\end{equation}

\subsubsection{Trojan horse attacks}
THAs have been recognized as powerful eavesdropping strategies ~\cite{Lucamarini2015BoundTHA,Zhang2021Securingsys}. Here, we consider a special THA in which Eve can inject strong light into Alice's and Bob's sites through the quantum channel and obtain the phase modulation information by measuring the light reflected from Alice's and Bob's optical devices. Here, we can assume that the back-reflected strong light Eve sends to Alice's (Bob's) site is a strong coherent pulse:
\begin{equation}
	\Ket{\zeta_{\rm T}}_{\rm E}=e^{-\mu/2}\Ket{v_{\rm T}}_{\rm E}+\sqrt{1-e^{-\mu}}\Ket{\xi_{\rm T}}_{\rm E},
\end{equation}
where $\rm T \in \{{\rm 0_{X},0_{Y},1_{X},1_{Y}}\}$ denotes the bit and basis choices of Alice and Bob and $\mu$ is the intensity of Eve's back-reflected light. Here, we apply the model of THAs, in which $\Ket{\zeta_{\rm T}}_{\rm E}$ is a coherent state and $\Ket{v_{\rm T}}_{\rm E}$ is the vacuum state~\cite{pereira2019quantum}. For simplicity, we consider the worst scenario in which the overlap is $\fid{\xi_{\rm T}}{\xi_{{\rm T}^{\prime}}}_{\rm E}=0$. When jointly considering the SPFs and side-channel effects, the basis-dependent states can be written as
\begin{equation}
	\begin{aligned}
		&\Ket{\Psi_{\rm X}}^{\prime\prime}=\frac{1}{\sqrt{2}}\left( \ket{0_{\rm X}}\ket{\alpha_0^{\prime}}\Ket{\zeta_{\rm 0X}}_{\rm E}+\ket{1_{\rm X}}\ket{\alpha_1^{\prime}}\Ket{\zeta_{\rm 1X}}_{\rm E} \right),\\
		&\Ket{\Psi_{\rm Y}}^{\prime\prime}=\frac{1}{\sqrt{2}}\left( \ket{1_{\rm Y}}\ket{\alpha_2^{\prime}}\Ket{\zeta_{\rm 1Y}}_{\rm E}+\ket{0_{\rm Y}}\ket{\alpha_3^{\prime}}\Ket{\zeta_{\rm 0Y}}_{\rm E} \right).\\	
	\end{aligned}
\end{equation}

\subsubsection{{Pulse} correlations}
{Pulse} correlation is another security loophole in a practical QKD system~\cite{Yoshino2018,Mizutani2019,Pereiraeaaz4487,zapatero2021security,PhysRevA.104.062611}, which is caused by different modulations between coherent pulses and can reveal the modulation information to Eve. Because vacuum state modulation is avoided in this study, it is not necessary to consider the intensity correlation between pulses~\cite{zapatero2021security}. However, since our protocol adopts a phase encoding scheme, phase modulation also introduces {pulse} correlations, which can be observed in our experiments. Fortunately, it has been theoretically demonstrated that {classical pulse} correlations can be interpreted as side channels~\cite{Pereiraeaaz4487}. {Here we also discuss classical pulse correlations and leave characterization of quantum pulse correlations to be the future study. Furthermore, based on the recent study in Ref.~\cite{PhysRevA.104.062611}, we firstly theoretically present characterization of long-range pulse correlations where the length of correlations is set as $l_c$. Ignoring other source flaws, we express the entangled state of $N$ pulses sent by Alice as
	\begin{equation}
		\Ket{\Psi}_{\rm A}=\frac{1}{M}\sum_{j_n}\sum_{j_{n-1}}...\sum_{j_1}\bigotimes_{k=1}^{N}\sqrt{p_t}\Ket{j_k}_{A_k}\Ket{\psi_{j_k|j_{k-1}}}_{C_k},	 
	\end{equation}
	where $1/M$ is a coefficient of normalization and $j_k\in\{0_{\rm X}, 1_{\rm X}, 0_{\rm Y}, 1_{\rm Y}\}_{k=1,2,...,N}$ denotes the qubit state representing bit and basis choices of Alice. $p_t \in \{p_x,p_y\}$ is related to basis choices since Alice biasedly chooses states in the X and Y bases. $\Ket{\psi_{j_k|j_{k-1}}}_{C_k}$ is the $k$th state Alice send to Charlie which is affected by long-range pulse correlations. The conclusion in Refs.~\cite{Pereiraeaaz4487,PhysRevA.104.062611} shows that the actual $k$th state can also be expressed as
	\begin{equation}
		\Ket{\psi^{\rm act}_{j_k|j_{k-1}}}_{{\rm A}_{k}}=a_{k}\Ket{\Phi}_{{\rm A}_{k}}+\sqrt{1-a^2_k}\Ket{\Phi^{\perp}}_{{\rm A}_k}.
	\end{equation}
	$\Ket{\Phi}_{{\rm A}_{k}}$ represents the ideal state which is side channel free and $\Ket{\Phi^{\perp}}_{{\rm A}_k}$ is the state with side channels, which is orthogonal to $\Ket{\Phi}_{{\rm A}_{k}}$. There is no need to derive the specific expression of $\Ket{\Phi^{\perp}}_{{\rm A}_k}$. To minimize $a_k$ means we maximize the pulse correlations which is considered as side channels. We directly derive the relation between the lower bound of $a_k$ and parameters characterizing pulse correlations of different lengths through outcomes in Ref.~\cite{PhysRevA.104.062611} as $a_k=\prod_{d=1}^{l_c}\sqrt{1-\epsilon_{d}}$. $\epsilon_d$ is the parameter characterizing influence of pulse correlations when the length of correlation is $d$. For simplicity, we introduce one parameter $\epsilon$ to generalize the formula of $a_k$, which makes the relation $a_k=\sqrt{1-\epsilon}$. For different basis choices of the X and Y bases, the relation remains. Therefore,
	we can obtain the state-dependent states as
	\begin{equation}\label{finalstate}
		\begin{aligned}
			&\Ket{\Psi_{\rm X}}=\sqrt{1-\epsilon}\Ket{\Psi_{\rm X}}^{\prime\prime}+\sqrt{\epsilon}\Ket{\Psi_{\rm X}^{\perp}},\\
			&\Ket{\Psi_{\rm Y}}=\sqrt{1-\epsilon}\Ket{\Psi_{\rm Y}}^{\prime\prime}+\sqrt{\epsilon}\Ket{\Psi_{\rm Y}^{\perp}},\\
		\end{aligned}
	\end{equation}
	removing subscripts $j_k$ and introducing subscripts of basis choices to accomodate the GLLP analysis. Here we also consider the worst situation for the senders, where $\fid{\Psi_{\rm X}^{\perp}}{\Psi_{\rm Y}^{\perp}}=0$. In the experiment, we could measure the deviation of phase modulations caused by pulse correlations, which is called the pattern effect~\cite{Yoshino2018}.} Here, we only recorded the maximum phase deviation of nearest-neighbor pulse correlations, which is denoted as $\psi$. {We remark that in our protocol, pulse correlations of different lengths can also be recorded through our method measuring pattern effects.} The relation between $\epsilon$ and $\psi$ can be built through the following analysis. For simplicity, our discussion is also based on the condition that imperfections other than the {pulse} correlations are neglected. For example, when the ideal state without any imperfection is $\Ket{-\alpha}$, corresponding to bit $1$ on the X basis, the phase modulations of this state will be influenced by long-range classical {pulse} correlations owing to the phase modulation of other pulses. The practical state corresponds to $\Ket{-e^{i\psi}\alpha}$. Similar to the basic idea for finding the reference state, we can write the practical state as
\begin{equation}
	\Ket{-e^{i\psi}\alpha}=e^{(e^{i\psi}-1)|\alpha|^2}\Ket{-\alpha}
	+\sqrt{1-e^{2(e^{i\psi}-1)|\alpha|^2}}\Ket{-\alpha^{\perp}},
\end{equation}
where $\fid{-\alpha}{-e^{i\psi}\alpha}=e^{(e^{i\psi}-1)|\alpha|^2}$ and there is no need to write down the specific expression of $\Ket{-\alpha^{\perp}}$. The definition of $\epsilon$ leads to the formula $\sqrt{1-\epsilon}=e^{(e^{i\psi}-1)|\alpha|^2}$, and we obtain the relation $\epsilon=1-e^{|\alpha|^2(2{\rm cos}\psi-2)}$. For other bit and basis choices, it is readily observed that the formula remains the same. In the experimental implementation, we will measure the pattern effect whose values can be seen in Supplementary Material. The pattern effect can be expressed by $\psi$ in the form of ${\rm sin}\psi$, which means $\psi$ can be derived through measurements of pattern effects. To be precise, for state $\Ket{\alpha}$, after phase modulation, the state can be expressed as $\Ket{\alpha'}=e^{i(\psi+\theta)}\Ket\alpha$. Consequently, the interference result of $\Ket{\alpha'}$ and $\Ket\alpha$ is $(e^{i\psi}\pm1)\Ket{\alpha}/{\sqrt2}$. Thus, the deviation of the pulse intensity after interference is ${\rm sin}\psi$.

\begin{figure*}[t]
	\includegraphics[width=17.2cm]{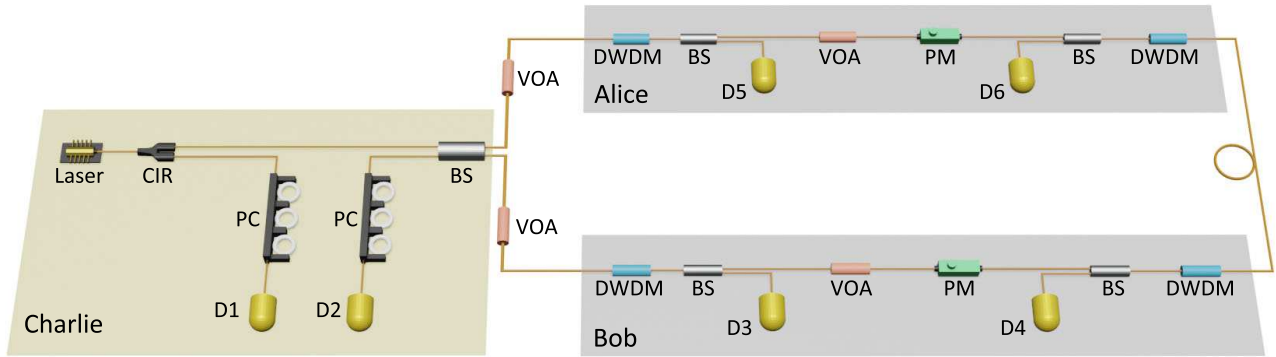}
	\caption{Experimental setup of our plug-and-play system. Optical pulses are directly generated by a pulsed laser source and then pass through a circulator (CIR). These pulses are separated into two pulse sequences with different propagation directions using a $50:50$ beam splitter (BS). The clockwise pulse sequence is modulated by a phase modulator (PM) controlled by an arbitrary waveform generator after passing dense wavelength division multiplexing (DWDM) and a BS. A counterclockwise pulse sequence undergoes a similar process. Finally, these two pulses interfere in the BS and are detected by two superconducting nanowire single-photon detectors D1 and D2. Polarization controllers (PC) are used to modify the polarization of the incident pulses to achieve maximum detection efficiencies.}
	\label{fig03}
\end{figure*}	

\subsubsection{The calculation of the basis-dependent states}
Here, we need to calculate the analytic formula for the bounded fidelity to  estimate the imbalance $\Delta$. The fidelity $\fid{\Psi_{\rm X}}{\Psi_{\rm Y}}$ becomes $\fid{\Psi_{\rm X}}{\Psi_{\rm Y}}=(1-\epsilon)\fid{\Psi_{\rm X}}{\Psi_{\rm Y}}^{\prime\prime}$ and
\begin{equation}
	\begin{aligned}
		\fid{\Psi_{\rm X}}{\Psi_{\rm Y}}^{\prime\prime}=&\frac{1}{4}e^{-\mu}{\rm cos}^2\theta\lbrack(1-i)\fid{\alpha_0}{ie^{i\delta_2}\alpha_2}\\
		&+(1-i)\fid{-e^{i\delta_1}\alpha_1}{-ie^{i\delta_3}\alpha_3} \\
		&+(1+i)\fid{\alpha_0}{-ie^{i\delta_3}\alpha_3} \\
		&+(1+i)\fid{-e^{i\delta_1}\alpha_1}{ie^{i\delta_2}\alpha_2})\rbrack,
	\end{aligned}
\end{equation}
which can be applied to the calculation of joint fidelity $|\fid{\Psi_{\rm Y}}{\Psi_{\rm X}}|^2$. For simplicity, we further assumed that $\delta_1=\delta_2=\delta_3=\delta$ and $\alpha_0=\alpha_1=\alpha_2=\alpha_3=\alpha^{\prime}$, which represent the maximum deviations of the intensity and phase modulations, respectively, as shown in Eq.~\eqref{practicalfinal}.

\subsection{Experimental implementation}\label{sec:introduction}

Here, we present a proof-of-principle experimental demonstration of the FP-MDIQKD protocol. The setup is illustrated in Fig.~\ref{fig03}. This is a plug-and-play construction consisting of a Sagnac interferometer, similar to the system in Refs.~\cite{PhysRevLett.123.100506,yin2019measurement}. We use the Sagnac loop to stabilize the phase fluctuation of the channel automatically, and all optical fibers maintain polarization.

Optical pulses are generated using a homemade pulsed laser. Pulses with a frequency of 100 MHz have an extinction ratio greater than 30 dB and a pulse width of less than 300 ps. Charlie sends the optical pulses to Alice and Bob. These optical pulses pass through a circulator and are then separated by a 50:50 BS into two identical pulses. The two pulses are phase-modulated by Alice and Bob. Specifically, clockwise (counterclockwise) pulses are modulated by Alice (Bob). In addition, four DWDMs are used as bandpass filters, and four BSs are utilized to monitor the intensity of pulses entering Alice's or Bob's site, with the aim of protecting against attacks from injected pulses. In fact, we did not monitor the intensity and filter pulses during implementation because of resource limitations, which should be addressed for security. However, these processes can be directly added to our system  with little influence on the results. Alice (Bob) chooses the X basis with a probability of $p_x=90\%$ and the Y basis with a probability of $p_y=10\%$. Note that random numbers are generated by Python's module random and the cycle length is 10000. The radio frequency signals for modulation and clock signals of the experimental system are produced by an arbitrary waveform generator with a sampling rate of 2.5 Gs/s (Tabor Electronics, P2588B).

After the phase modulations, Alice (Bob) send pulses through a variable optical attenuator before the Charlie's beam splitter.  The loss between Alice (Bob) and Charlie is adjusted to simulate the loss of the communication channel. Pulses from Alice and Bob interfere at the BS. One output of this BS is detected by a single-photon detector D1 via the circulator, whereas the other output is followed directly by D2. {The experimental parameters include the laser wavelength $\lambda=1550.14$ nm, total efficiencies of the optical elements at the detection site, $\eta_{\rm det1}=75.3\%$ and $\eta_{\rm det2}=86.1\%$, and detection efficiencies of the two detectors, $\eta_{\rm D1}=88.0\%$ and $\eta_{\rm D2}=85.5\%$. $\eta_{\rm det1}$ consists of the efficiencies of three main components: the beam splitter, circulator, and polarization controller, whereas $\eta_{\rm det2}$ includes the beam splitter and polarization controller. Generally, for $D_1$, the total detection efficiency $\eta_1$ (considering the loss in Charlie) is $66.3\%$ and the dark count rate $p_{d,1}$ is $1.6\times 10^{-8}$ (1.5 ns per time window). For $D_2$, the detection efficiency $\eta_2$ is $73.6\%$ and the dark count rate $p_{d,2}$ is $2.5\times 10^{-8}$ (1.5 ns per time window).}	
	
We measured and quantified four types of source flaws in this system: imperfect intensity modulation, imperfect phase modulation, the extinction ratio of polarization (side channels), and pattern effects (pulse correlation). It should be noted that THAs cannot be quantified in this implementation, which is a flaw in the plug-and-play system, and we set them to $10^{-7}$~\cite{Lucamarini2015BoundTHA}. Imperfect intensity modulation is quantified by the optical power fluctuation $\xi$ of the pulses sent, because there is no intensity modulation in this protocol. Imperfect phase modulation can be quantitively characterized by the deviation $\delta$ between the actual and expected phases. For side channels $\theta$, we measured the the extinction ratio of polarization ${\rm tan}\theta$ of the pulses when the system was running. The pattern effects were quantified by the interference results of the same phases under different patterns and 0 phase, which is denoted as $\psi$. All parameters corresponding to those are introduced in Eq.~\eqref{practicalfinal}. The results are presented as follows. The five parameters related to the imperfection of
realistic sources included the optical power
fluctuation $\xi=1.11\%$, the extinction ratio of polarization tan$\theta=10^{-2.65}$, phase
shift $\delta=0.062$, pattern effect $\psi=8.54\times10^{-3}$, and THAs $\mu=10^{-7}$. Detailed descriptions of the derivation of these parameters are provided in the Supplementary Material.

\section{Results}

\begin{figure*}[t]
	\centering
	\includegraphics[width=2\columnwidth]{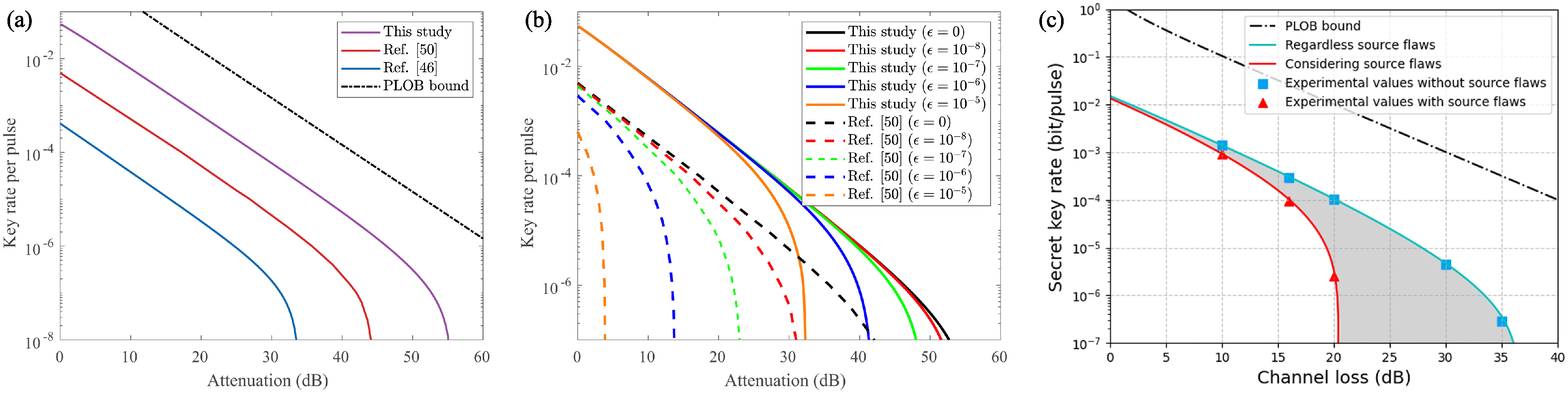}\\
	\caption{Key rate vs. the overall transmission loss. (a) When there are no source flaws, we compare the performance of our protocol (violet line) with that of two other protocols in Ref.~\cite{PhysRevApplied.15.034072} (red line) and Ref.~\cite{PhysRevApplied.12.054034} (blue line) and the PLOB bound~\cite{PhysRevLett.102.050503,Pirandola2017} (black dotted line). An evident improvement in the key rate indicates that our protocol outperformed the other two protocols. (b) We applied the same parameter $\epsilon$ as in Ref.~\cite{PhysRevApplied.15.034072} to characterize general side channels and depict the performance of secret key rates in both our protocol and the protocol in Ref.~\cite{PhysRevApplied.15.034072} for different values of $\epsilon$ ($0,~ 10^{-8},~10^{-7},~10^{-6},~10^{-5}$) using different colors. Evidently, our protocol (solid lines) outperforms the protocol described in Ref.~\cite{PhysRevApplied.15.034072} (dashed lines) with respect to key rate. With $\epsilon=10^{-5}$, the orange solid line can still generate a secure key rate when the total loss is 30 dB, whereas the orange dashed line can barely generate a secure key rate. {(c) Considering the experimental results, we depicted the secret key rate when the total number of pulses sent by Alice or Bob $N=10^{10}$. The filled symbols indicate the experimental results obtained for whether source flaws were considered . The solid lines correspond to the theoretical simulation results for our protocol, under different assumptions.}}\label{fig02}
\end{figure*}

\subsection{Simulation results}
To evaluate the performance of our protocol theoretically and experimentally, we simulate the secure key rates of our protocol in two different scenarios, sharing one precondition that we only conduct simulations in the asymptotic scenario. Specifically, we set the detector efficiency $\eta_{\rm d}=1$ and the dark count rate $p_{\rm d}=10^{-8}$. The error correction efficiency $f$ was set to 1.16. The simulation results are shown in Fig.~\ref{fig02}, respectively.

In the first scenario, shown in Fig.~\ref{fig02}a, we only considered the ideal scenario with no source flaws, which means that the parameters characterizing the source flaws can easily be set to zero. For comparison, we also depict the key rates of the two other side-channel-free protocols in Refs.~\cite{PhysRevApplied.15.034072}  and ~\cite{PhysRevApplied.12.054034}. The simulation results clearly illustrate that, when the total attenuation approaches 50 dB, only our protocol can generate secret keys. The final key rate of our protocol cannot break the PLOB bound, which results from the pessimistic assumption made in the GLLP analysis, where Eve is assumed to be able to enhance the fidelity decay with the channel losses. For improving the performance of our final key rate, the decoy state method~\cite{PhysRevLett.91.057901,PhysRevLett.94.230503,PhysRevLett.94.230504} can be applied, which is the key to overcome the PLOB bound in twin-field-type QKD protocols~\cite{tamaki2018information}. However, with the same transmission loss, the key rate of our protocol was approximately one order of magnitude higher than that of the protocol in Ref.~\cite{PhysRevApplied.15.034072}, and approximately two orders of magnitude higher than that of the protocol in Ref.~\cite{PhysRevApplied.12.054034}.

In the second scenario, as illustrated in Fig.~\ref{fig02}b, we analyzed the security of our protocol compared to that in Ref.~\cite{PhysRevApplied.15.034072},  which generalizes all the side channels to a single parameter, $\epsilon$. From the formula for the practical secure key rate in Eq.~\eqref{finalstate}, if we only consider the pulse correlation flaws and set the other parameters to zero, $\epsilon$ can be converted into the parameter introduced in Ref.~\cite{PhysRevApplied.15.034072} , which characterizes general side channels. Based on this, we evaluated the key rate of our protocol for different degrees of general side channels ($\epsilon=0, ~ 10^{-8},~10^{-7},~10^{-6},~10^{-5}$). In Fig.~\ref{fig02}b, when $\epsilon$ increases, a decrease in the key rate can be observed both in our study and in the protocol in Ref.~\cite{PhysRevApplied.15.034072}, and a better performance can be observed in our secret key rate with the same $\epsilon$. The performance of our key rate means the decay of the fidelity causes the lower key rate and the decay of the fidelity means the more deviation between the practical states and the ideal states, which is why we consider the worst case of the fidelity in our theoretical analysis and experimental implementation. Additionally, our protocol can still generate a considerable secret key rate when $\epsilon=10^{-5}$.

\subsection{Experimental results}
To demonstrate the advantages of our protocol when considering source flaws, we implemented experiments under three values of channel losses between Alice and Bob: 10, 16, and 20 dB, with the number of pulses sent by Alice (Bob) $N=10^{10}$, as shown in Fig.~\ref{fig02}c. We further performed experiments under channel losses of 30 and 35 dB, with the assumption that the sources are perfect. The shaded area between the blue and red lines shown in Fig.~\ref{fig02}c is unsafe and need to be considered seriously. The blue line can be reached by rigorously narrowing the parameters in Eq.~\eqref{practicalfinal} to approximately 0. Considering imperfect sources, our protocol can obtain secure key bits when the channel loss is over 20 dB, which means that it can be implemented over 100 km using existing technologies. The key rate of 253 bps we achieved under a channel loss of 20 dB is the best result for the transmission distance and key rate considering both source and detection imperfections. Under 10 dB ($\sim$50 km), the key rate of our protocol is over 90 kbps, whereas under 16 dB ($\sim$80 km), the key rate is over 9 kbps. Regardless of the source flaws, we achieved a key rate of 29.2 bps with a channel loss of 35 dB. Detailed experimental results are presented in Table~\ref{tab04}. Although we cannot actually demonstrate THAs in our experiment, a recent study focused on resisting THAs, and its creative design of optical power limiters can also be applied in our study against THAs~\cite{Zhang2021Securingsys}. In addition, despite the fact that optical fibers were not used in the actual implementation and the two users were not separated, as a proof-of-principle demonstration, we focused on the feasibility of our protocol rather than aiming to establish a complete system with all necessary hardware. Combined with the developed phase-locking and phase-tracking technology in the implementation of twin-field QKD~\cite{Yin2016MDI404,Fang2020impltf,Pittaluga2021,Chen2021,Wang2022}, our protocol containing two independent users can be directly implemented in real optical fibers, which is illustrated in detail in the Supplementary Material.

\begin{table*}
	\center
	\caption{Summary of experimental data. We tested the key rates for different channel losses. The table shows the number of pulses sent, the expected intensity of pulses $\mu^{\prime}=|\alpha|^2$,  the number of detection events under corresponding bases, the experimental quantum bit error rates under the X basis and Y basis, the number of clicks under the X basis $S_{\rm X}^{\rm L}$, the key rate $R^*$ regardless of source flaws, and the key rate $R$ considering source flaws.}
	\setlength{\tabcolsep}{0.485cm}
	\begin{tabular}
		{cccccccc}  \hline  \xrowht[()]{7pt}
		Loss (dB) & $N$ &$\mu^{\prime}$ & $E_{\rm b}^{\rm X}$ $(\%)$ & $E_{\rm b}^{\rm Y}$ $(\%)$& $S_{\rm X}^{\rm L}$& $R^*$& $R$ \\ \hline\xrowht[()]{7pt}
		10  &$10^{10}$& 1.38$\times 10^{-2}$ & 0.23 &0.36 & 49398692& 1.41$\times 10^{-3}$& 9.10$\times 10^{-4}$ \\  \xrowht[()]{7pt}
		16  & $10^{10}$& 6.40$\times 10^{-3}$ &0.23 &0.38 &11421107 & 3.00$\times 10^{-4}$&9.39$\times 10^{-5}$\\ \xrowht[()]{7pt}
		20  & $10^{10}$& 3.30$\times 10^{-3}$ &0.22 &0.36 &3718173 & {1.02}$\times 10^{-4}$&2.53$\times 10^{-6}$\\ \xrowht[()]{7pt}
		30  &$10^{10}$&  1.31$\times 10^{-3}$ &0.27&0.42  &464991  &{4.40$\times 10^{-6}$}& 0\\ \xrowht[()]{7pt}
		35  &$10^{10}$&  6.27$\times 10^{-4}$&0.33 &0.57  &124117 &{2.92$\times 10^{-7}$}& 0 \\ \hline
	\end{tabular}
	
	\label{tab04}
\end{table*}
	
\section{Discussion and conclusion}
Based on the fact that tremendous progress has been witnessed in the deployment of QKD, practical secure QKD protocols are becoming an important research direction in quantum communication since they are relevant for the practical deployment of QKD. In this study, we adopt the RT method to guarantee security of an FP-MDIQKD protocol using coherent states against realistic flawed sources. We further demonstrate the feasibility of the protocol through a proof-of-principle experimental implementation~\cite{Lucamarini2018}. For source flaws, we consider SPFs, side channels caused by mode dependencies, THAs, and {classical pulse} correlations, which typically cover all possible source imperfections that are currently observed in experimental implementations. Although we only consider the special polarization side channel and one special THA, we note that other similar side channels and THAs can also be included in the calculation of fidelity with extra parameters. However, actually, imperfections in side channel space, which are not considered in our state representation and our calculation of fidelity, also have influences on the fidelity and the experimental results, which is a remaining problem for future study. In addition, we offer relations between parameters characterizing source flaws and experimental data, and provide specific implementations in our experiment on how to define and measure these parameters, which can also be compatible to other types of potential source flaws, which means higher practicality of our study in experimental and real-life implementations. In addition, the simulation results show that our protocol generates a higher secure key rate against source flaws than the other QKD protocols in Refs.~\cite{PhysRevApplied.15.034072} and ~\cite{PhysRevApplied.12.054034}). For experimental implementation, we also present a finite-key analysis against coherent attacks.

Furthermore, we experimentally demonstrate our protocol using a plug-and-play system. With perfect sources, our protocol can be demonstrated over 35 dB , which is a considerable improvement in the transmission distance in the realm of practical secure QKD. By removing assumptions of the perfect source and detection site, our protocol can still achieve a key rate of 253 bps with a 20 dB channel loss, which outperforms other experimental implementations of practical secure QKD protocols. Besides, we also achieve key rates of 91.0 and 9.39 kbps at 10 and 16 dB, respectively. The transmission distance of our protocol was constrained using five parameters relevant to imperfect sources. However, these five parameters can be optimized. Thus, the key rate and transmission distance of this protocol can be further improved. First, the measurements of these parameters are limited by the accuracy and precision of the devices. For example, the accuracy of the power meter measuring optical power fluctuations is limited, and the power meter itself has jitters, which may result in larger fluctuations. Additionally, the values of these parameters can be further reduced using available technologies. More accurate phase modulation can be achieved by using more phase modulators, which can avoid nonlinear amplification of radio frequency signals. The extinction ratio of polarization can be further improved by fusing optical fibers or by adding a polarizer at the user's site. With developed technologies, these parameters will be closer to 0. Note that our scheme does not need to resort to immature technologies such as entanglement or single-photon sources, which makes our scheme more practical. Overall, our study further proves the feasibility of secure QKD with arbitrary flawed devices, and holds the potential to be applied to the future deployment of practical secure QKD.

\section*{Conflict of interest}
	The authors declare that they have no conflict of interest.
	
\section*{Acknowledgments}
	
	This study was supported by the Natural Science Foundation of Jiangsu Province (BK20211145), the Fundamental Research Funds for the Central Universities (020414380182), the Key Research and Development Program of Nanjing Jiangbei New Aera (ZDYD20210101), the Program for Innovative Talents, and Entrepreneurs in Jiangsu (JSSCRC2021484). We thank P. Liu and Y.-S. Lu for valuable discussion and
	Y. Bao and S.-C. Huang for experimental assistance.

\section*{Author contributions}
Hua-Lei Yin and Zeng-Bing Chen supervised the study. Jie Gu and Hua-Lei Yin built the theoretical model. Xiao-Yu Cao, Yao Fu, Zong-Wu He, Ze-Jie Yin, and Hua-Lei Yin designed and performed the experiment. Jie Gu, Xiao-Yu Cao, Yao Fu, and Hua-Lei Yin analyzed the experimental data. Jie Gu, Xiao-Yu Cao, and Hua-Lei Yin co-wrote the manuscript with input from the other authors. All authors have discussed the results and proofread the manuscript.

\appendix

\section{Formula of the final key rate}
In our four-phase measurement-device-independent QKD protocol, the classical bits can be encoded through phase differences of coherent pulses from Alice and Bob and the setting of the X and Y bases introduces the idea of "quantum coin" where we can introduce $\Delta$ as the imbalance of a quantum coin to bound the information leakage. The imbalance $\Delta$ can be calculated through the bounded fidelity discussed in the GLLP security proof~\cite{lo2007security,Koashi_2009}, which can be naturally extended to the security analysis of the reference technique~\cite{Pereiraeaaz4487}. In the ideal scenario with no source flaws, Alice's entangled basis-dependent states are expressed as
\begin{equation}\tag{S1}
	\begin{aligned}
		&\Ket{\Psi_{\rm X}}_{\rm Aa}=\frac{1}{\sqrt{2}}(\ket{0_{\rm X}}_{\rm A}\ket{\alpha}_{\rm a}+\ket{1_{\rm X}}_{\rm A}\ket{-\alpha}_{\rm a}),\\
		&\Ket{\Psi_{\rm Y}}_{\rm Aa}=\frac{1}{\sqrt{2}}(\ket{1_{\rm Y}}_{\rm A}\ket{i\alpha}_{\rm a}+\ket{0_{\rm Y}}_{\rm A}\ket{-i\alpha}_{\rm a}),
	\end{aligned}
\end{equation}
where $\{\ket{0_{\rm X}}, \ket{1_{\rm X}}\}$ are eigenstates of the Pauli operator $\sigma_x$ and $\{\ket{0_{\rm Y}}, \ket{1_{\rm Y}}\}$ are eigenstates of the Pauli operator $\sigma_y$. The same equations of Bob's entangled basis-dependent states can be obtained in the similar expressions with the system B denoting the virtual qubit space and the system b representing the actual space of coherent pulses. Here, the basis-dependence of Alice's (Bob's) entangled states can be related to an equivalent virtual protocol considering the balance of a quantum coin~\cite{lo2007security}, where Alice (Bob) measures the coin in the Z basis to determine the basis choice. Here the basis-dependent states can be expressed as $\Ket{\Psi_{\rm X}}$ and $\Ket{\Psi_{\rm Y}}$ without distinguishing the sender. The joint state can be taken to be
\begin{equation}\tag{S2}
	\begin{aligned}
		\Ket{\Phi}=\sqrt{p_x}\ket{0_{\rm Z}}\Ket{\Psi_{\rm X}}+\sqrt{p_y}\ket{1_{\rm Z}}\Ket{\Psi_{\rm Y}},
	\end{aligned}
\end{equation}
where $p_x$ and $p_y$ separately represent the probabilities of sending states in the X basis and the Y basis. Note that we assume the measurement of the quantum coin will be delayed after Charlie has finished her measurement and eavesdropping.	
The final key rate of this protocol is 
\begin{equation}\tag{S3}
	R=Q^{x}[1-fH(E_{\rm b}^{x})-H(E_{\rm p})],
\end{equation}
where $Q^{x}=(1-p_{\rm d})[1-(1-2p_{\rm d})e^{-2\eta|\alpha|^2}]$ and $E_{\rm b}^{x}Q^{x}=e_{\rm d_{x}}(1-p_{\rm d})[1-(1-p_{\rm d})e^{-2\eta|\alpha|^2}]+(1-e_{\rm d_{x}})p_{\rm d}(1-p_{\rm d})e^{-2\eta|\alpha|^2}$. $Q^{x}$ is the gain in the X basis and $E_{\rm b}^{x}$ means the bit error rate under the X basis. $e_{\rm d_{x}}$ means the misalignment error rate in the X basis of the four-phase measurement-device-independent QKD system. The total efficiency is $\eta$ and we use $|\alpha|^2$ to express the intensity of weak coherent states. $E_{\rm p}$ represents the phase error rate under the X basis, which is bounded by $\Delta$ and the bit error rate in the Y basis $E_{\rm b}^{y}$. The equation of $E_{\rm b}^{y}$ is similar to that of $E_{\rm b}^{x}$ with its separate misalignment error rate $e_{\rm d_{y}}$ that $E_{\rm b}^{y}Q=e_{\rm d_{y}}(1-p_{\rm d})[1-(1-p_{\rm d})e^{-2\eta|\alpha|^2}]+(1-e_{\rm d_{y}})p_{\rm d}(1-p_{\rm d})e^{-2\eta|\alpha|^2}$. Here for simulation, we just consider the situation that $E_{\rm b}^{y}$ is equal to $E_{\rm b}^{x}$ since we set the misalignment error rates to be 0 without the finite-key analysis. Note that $E_{\rm b}^{y}$ is not rigorously equal to $E_{\rm b}^{x}$ in the practical scenario since $e_{\rm d_{y}}$ evidently differs from $e_{\rm d_{x}}$ in experiment. $\eta=\eta_{\rm d}\times 10^{-L/20}$, where $L$ $({\rm dB})$ is the total attenuation of the channel.\\
Based on the fact that our protocol considers an biased basis choice, we provide the relation between the phase error rate in the X basis $E_{\rm p}$ and the bit error rate in the Y basis $E_{\rm b}^{y}$. According to the procedure of our protocol, the simplified entangled state shared by Alice and Bob in two bases can be expressed as 
\begin{equation}\tag{S4}
	\begin{aligned}
		&\Ket{X}=\Ket{\Psi_{\rm X}}_{\rm Aa}\Ket{\Psi_{\rm X}}_{\rm Bb},\\
		&\Ket{Y}=\Ket{\Psi_{\rm Y}}_{\rm Aa}\Ket{\Psi_{\rm Y}}_{\rm Bb},\\
	\end{aligned}
\end{equation}
where we denote regardless of the measurement result of Charlie. We introduce the error operator by Alice and Bob, $E_{\rm AB}^{y}=\Ket{0_{\rm Y}1_{\rm Y}}_{\rm AB}\Bra{0_{\rm Y}1_{\rm Y}}+\Ket{1_{\rm Y}0_{\rm Y}}_{\rm AB}\Bra{1_{\rm Y}0_{\rm Y}}$. The phase error rate of the $X$ basis and the bit error rate of the Y basis
are defined as 
\begin{equation}\tag{S5}
	\begin{aligned}
		E_{\rm p}&=\Bra{X}M_{ab}^{\dag}E_{\rm AB}^{y}M_{ab}\Ket{X},\\
		E_{\rm b}^{y}&=\Bra{Y}M_{ab}^{\dag}E_{\rm AB}^{y}M_{ab}\Ket{Y}, \\
	\end{aligned}
\end{equation}
where $M_{ab}$ is the Kraus operator corresponding to the announcement of Charlie. Note that the  normalization has been assumed. Similar with the method in Ref.~\cite{PhysRevA.104.022423}, we have
\begin{equation}\tag{S6}
	\begin{aligned}
		E_{\rm p}=&|\fid{X}{Y}|^2\Bra{Y}\mathbf{E}\Ket{Y}+|\fid{X}{Y^{\bot}}|^2\Bra{Y^{\bot}}\mathbf{E}\Ket{Y^{\bot}}\\
		&+\fid{X}{Y}\fid{X}{Y^{\bot}}(\Bra{Y}\mathbf{E}\Ket{Y^{\bot}}+\Bra{Y^{\bot}}\mathbf{E}\Ket{Y}),\\
	\end{aligned}
\end{equation}
where we let $\mathbf{E}=M_{ab}^{\dag}E_{\rm AB}^{y}M_{ab}$ and use the fact that $I=\ket{Y}\bra{Y}+\ket{Y^{\bot}}\bra{Y^{\bot}}$.
The fidelity of quantum states can be expressed as $1-2\Delta' =\fid{X}{Y}=|\fid{\Psi_{\rm X}}{\Psi_{\rm Y}}|^2$. 
Eve can enhance the source flaws by exploiting the loss of channel, which means that we should have the the worst situation $\Delta=\Delta'/Q$~\cite{lo2007security}.
Therefore, we can obtian the relation between $E_{\rm b}^{y}$ and $\Delta$,
\begin{equation}\tag{S7}
	1-2\Delta\leq\sqrt{E_{\rm b}^{y}E_{\rm p}}+\sqrt{(1-E_{\rm b}^{y})(1-E_{\rm p})}.
\end{equation}

\section{Finite-key analysis for the experimental implementation}
Here we provide the specific expression of the finite-key analysis for the experiment of our protocol. The number of final key bits $l$ has been expressed as
\begin{equation}\tag{S8}\label{S8}
	\begin{aligned}
		l=n_{x}\Bigg[1&-H(\overline{E}_{\rm p})-{\rm leak}_{\rm EC}\\
		&-\frac{1}{n_x}{\rm log}_2\frac{2}{\epsilon_{\rm c}}-\frac{1}{n_x}{\rm log}_2\frac{1}{4\epsilon^{2}_{\rm PA}}\Bigg],
	\end{aligned}
\end{equation}
which is $\epsilon_{\rm c}-{\rm correct}$ and $\epsilon_{\rm s}-{\rm secure}$, with $\epsilon_{\rm s}=\sqrt{\epsilon_{\rm F}}+\epsilon_{\rm PA}$~\cite{Lorenzo2021}. We directly offer $n_x$ through the experiment, together with $E_{\rm b}^{\rm X}$ in the formula of ${\rm leak}_{\rm EC}$. Based on Eq.~(S7), we have to consider the statistical fluctuation when calculating the bit error rate in the Y basis $E_{\rm b}^{\rm Y}$ and estimating the phase error rate considering eavesdroppers' coherent attacks. Here, we adopt the concentration inequality introduced in Ref.~\cite{kato2020concentration,Lorenzo2021} to guarantee security against coherent attacks when we bound the deviation between a sum of correlated random variables and its expected value.

Let $\xi_1,..., \xi_n$ be a sequence of Bernoulli random variables which are dependent and let $\Lambda_j=\sum_{u=1}^{j} \xi_u$. We denote $\mathcal{F}_j$ to be its natural filtration, which is the $\sigma$-algebra generated by $\{\xi_1,..., \xi_n\}$. Based on the conclusions in Ref.~\cite{kato2020concentration,Lorenzo2021}, we obtain that for any $b\geq 0$
\begin{equation}\tag{S9}
	\begin{aligned}
		&{\rm Pr}\left[\Lambda_n-\sum_{u=1}^{n}{\rm Pr}(\xi_u=1|\mathcal{F}_{u-1})\geq b\sqrt{n}\right]\leq {\rm exp}[-2b^2],\\
		&{\rm Pr}\left[\sum_{u=1}^{n}{\rm Pr}(\xi_u=1|\mathcal{F}_{u-1})\geq b\sqrt{n}-\Lambda_n\right]\leq {\rm exp}[-2b^2].\\
	\end{aligned}
\end{equation}
By equating the right hand of Eq.~(S9) to be $\epsilon_{\rm F}$ and solving for $b$, we have that
\begin{equation}\tag{S10}
	\begin{aligned}
		&\sum_{u=1}^{n}{\rm Pr}(\xi_u=1|\xi_1,..., \xi_{u-1})\leq \Lambda_n+\Delta_c,\\
		&\Lambda_n\leq \sum_{u=1}^{n}{\rm Pr}(\xi_u=1|\xi_1,..., \xi_{u-1})+\Delta_c,\\
	\end{aligned}
\end{equation}
where $\Delta_c=\sqrt{\frac{1}{2}n{\rm ln}\epsilon_{\rm F}^{-1}}$ and each of the bounds in Eq.~(S10) fail with probability at most $\epsilon_{\rm F}$. Eq.~(S10) is the simple bound of the concentration inequality but it is sufficiently tight for all opponents in our protocol. The concentration inequality~\cite{kato2020concentration} here is tighter than the widely applied Azuma’s inequality~\cite{Mizutani_2015} to guarantee security against coherent attacks when random variables are correlated and the encoding bases are not mutually unbiased~\cite{Lorenzo2021}. Then we will present how to derive the upper bound $\overline{E}_{\rm p}$. First, we set $n_y$ and $m_y$ to be the number of efficient outcomes and the number of bit errors in the Y basis, which is obtained through the  experiment. We derive an upper bound of the expectation value of the number of bit errors in the Y basis $m_y^{*}$. The upper bound of the expected bit error rate in the Y basis is ${E_{\rm b}^{y}}^{*}=m_y^{*}/n_y$, which can be used to calculate the upper bound of the expected value of the phase error rate in the X basis $E_{\rm p}^{*}$. Then the number of phase errors is obtained in the form of $m_{\rm p}^{*}=n_x E_{\rm p}^{*}$. We then estimate the upper bound of the observed number of phase errors in the X basis $\overline{m}_{\rm p}$ through the concentration inequality. Then $\overline{E}_{\rm p}$ can be naturally calculated with $\overline{E}_{\rm p}=\overline{m}_{\rm p}/n_x$. According to the experimental results, formulas in Eq.~(S10) is sufficiently tight for generating considerable final key rate in the finite-key regime so we just introduce the simple bound in Ref.~\cite{Lorenzo2021}. Without loss of generality, we set $\epsilon_{\rm c}=\epsilon_{\rm PA}=\epsilon_{\rm F}=10^{-10}$.

\section{Experimental details on derivation of parameters characterizing source flaws}
\subsection{Optical power fluctuation}
Imperfect intensity modulation can be quantitively characterized by the fluctuation of optical power over a period of time. {We test it by the optical power meter. Due to the limited range of power meter, we divide the fluctuation into two parts, the source fluctuation and the path fluctuation.} For the source part, we directly measured the optical power of the laser source with a 40 dB attenuation. The laser source is a homemade pulsed laser and Field Programmable Gate Array (FPGA) is used as the logic control of our laser source and to send laser drive signals. Since the circulator and beam splitters used in the system are fast-axis blocked, changes in polarization bring about fluctuations in intensity. Consequently, we measured the intensity of pulses sent by Alice with an appropriate attenuation to make the intensity around -70 dBm while the phase modulations are carried on. Besides, VOA also causes optical power fluctuation, so we add an additional attenuation each to the measurement of the fluctuation of path and source, which means the intensity of these pulses is lower than the minimum 
optical power (-103.7 dBm) we sent when considering source flaws. {Since the time of conducting one round of QKD is 100 seconds, we denote the difference between the maximum and minimum value of optical power in 100 seconds as the value of fluctuation.} As shown in Fig.~S1, the upper bound of optical power fluctuation in 1000 seconds is 1.11\%, which is the sum of source and path fluctuation.

\begin{figure}[ht!]
	\includegraphics[width=\columnwidth]{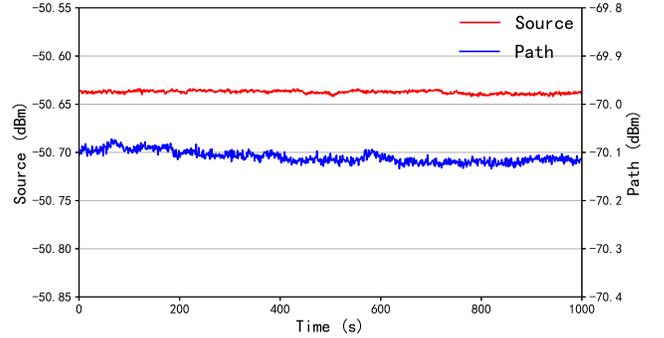}
	\caption{{Optical power fluctuation.} We divide the fluctuation into two parts, the source fluctuation and the path fluctuation. The optical power of pulses is measured once every 100 ms and we measure the fluctuation for 1000 seconds. Red line reflects the fluctuation of the source and the maximum difference within 100 s is 0.008 dB. While the blue line reflects the fluctuation of the path and the maximum value is 0.04 dB. Consequently, the total fluctuation is 0.048 dB ($1\pm 0.0111$). Consider the worst case, the optical power fluctuation is 1.11\%.}
	\label{fig05}
\end{figure}

\subsection{Phase shift}
We quantify the modulation error $\delta$ in the source through
calibrating Alice’s PM, in the plug-and-play QKD system. $\delta_\phi$ is defined as the difference between the actual phase and the expected phase $\phi$. In our protocol, we need to modulate 4 phases $\phi \in \{0, \pi/2, \pi, 3\pi/2\}$. Similar to Ref.~\cite{PhysRevA.92.032305}, only Alice performs phase modulation and Bob sets his own PM at a
fixed unmodulated phase 0. Note that the experimental setup here is the same as that used to demonstrate the QKD protocol. Alice and Bob are connected with a one-metre-long polarization maintained fiber. Alice scans the voltages applied to her PM according to the random phase selection in the experiment and the intensity of these pulses is 0.001. Charlie records the clicks for 100 seconds. Based on the phase selection, we can obtain the number of clicks for different phases and these counts are denoted by $D_1^{\phi}$ and $D_2^{\phi}$. For $D_1$, the total detection efficiency $\eta_1$ (considering the loss in Charlie) is $66.3\%$ and the dark count rate $p_{d,1}$ is $1.6\times 10^{-8}$. 
For $D_2$, the detection efficiency $\eta_2$ is $73.6\%$ and the dark count rate $p_{d,2}$ is $2.5\times 10^{-8}$. Since the dark count rates of two detectors are at the level of $10^{-8}$, dark counts have little influence on the computation and are ignored in the computation. The detailed experimental data can be seen in Table~S1. The function generator used is P25588B (Tabor Electronics) with 16-bit depth. The electrical drivers used are DR-VE-10-MO (iXblue) and PMs are PM-5S5-10-PFA-UV (Eospace). In our data analysis, we
use Hoeffding’s inequality~\cite{10.2307/2282952}. The upper bound of $\delta_\phi$ is given by Eq.~\eqref{10}. For $\phi \in \{\pi/2, \pi\}$, $\phi_0 = \phi$ and for $\phi =3\pi/2$, $\phi_0 = \phi - \pi$. Consequently, The error $\delta$ is upper bounded by the case of $\overline{\delta}_{\pi} = 0.062$.
Note that for statistical analysis, we only choose an appropriate time window without any other data processing, whose length is 2 ns.

\begin{table}[h!]
	\centering
	\caption{{Upper bounds of different phases.} The number of clicks detected by $D_1$ and $D_2$ under different phases and their upper bound $\overline{\delta}_\phi $.} 
	\setlength{\tabcolsep}{0.6cm} 
	\begin{tabular}
		{ccccc} \hline \hline \xrowht[()]{7pt}
		$\phi$ & $D_{1,\phi}$ & $D_{2,\phi}$ & $\overline{\delta}_\phi $\\ \hline
		0 & 5956406 &  1269 & -\\ 
		$\pi/2$& 343663 & 376778 & $0.012$ \\
		
		$\pi$ & 6627 & 6597672 & $0.062$\\
		
		$3\pi/2$ & 333491 & 367139  & $0.010$\\ \hline
		\hline
	\end{tabular}
	\label{tab05}
\end{table}

\begin{figure*}[htb]  
	\centering
	\begin{equation}\tag{S10}
		\overline{\delta}_\phi ={\rm max}\left\{ \left |\phi_0 - 2{\rm arctan}\sqrt{\frac{(\overline{D}_{2,\phi}^*-\underline{D}_{2,0}^*)/\eta_2} {(\underline{D}_{1,\phi}^*-\overline{D}_{2,0}^*)/\eta_1}}\right |, \left |\phi_0 - 2{\rm arctan}\sqrt{\frac{(\underline{D}_{2,\phi}^*-\overline{D}_{2,0}^*)/\eta_2} {(\overline{D}_{1,\phi}^*-\underline{D}_{2,0}^*)/\eta_1}}\right | \right\}. \label{10}
	\end{equation}
\end{figure*}

\subsection{Extinction ratio of polarization}
\begin{figure}[ht!]
	\includegraphics[width=\columnwidth]{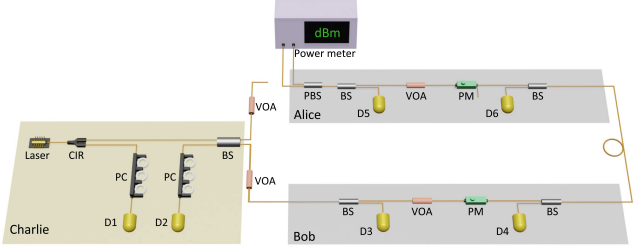}
	\caption{{The experimental setup for measuring the extinction ratio of polarization. The setup is similar to our plug-and-play system. A PBS (polarization beam splitter) is placed at the exit of a sending site. Two ports of the PBS are connected to the power meter and the optical power of fast-axis and slow axis can be obtained. Consequently, the relationship between the extinction ratio of pulse polarization and time can be acquired.}} 
	\label{figex}
\end{figure}
\noindent Since Eve could distinguish signals from
the polarization information of each state, we quantify this situation by measuring the extinction ratio of polarization in the plug-and-play system. We put a polarization beam splitter (PBS) at Alice's site to obtain the the extinction ratio of polarization of the pulses Alice sent, {which is the ratio of fast-axis optical power to slow-axis optical power. {The specific quantum states are shown as
		\begin{equation}\tag{S11}
			\begin{aligned}
				&\ket{\alpha_0^{\prime}}=\rm cos\theta_0\ket{\alpha_0}_{\rm H}+\rm sin\theta_0\ket{\alpha_0}_{\rm V},\\	
				&\ket{\alpha_1^{\prime}}=\rm cos\theta_1\ket{-e^{i\delta_1}\alpha_1}_{\rm H}+\rm sin\theta_1\ket{-e^{i\delta_1}\alpha_1}_{\rm V},\\
				&\ket{\alpha_2^{\prime}}=\rm cos\theta_2\ket{ie^{i\delta_2}\alpha_2}_{\rm H}+\rm sin\theta_2\ket{ie^{i\delta_2}\alpha_2}_{\rm V},\\
				&\ket{\alpha_3^{\prime}}=\rm cos\theta_3\ket{-ie^{i\delta_3}\alpha_3}_{\rm H}+\rm sin\theta_3\ket{-ie^{i\delta_3}\alpha_3}_{\rm V}.\\
			\end{aligned}
		\end{equation} Consequently, we denote the extinction ratio of polarization of pulses as ${\rm tan}\theta$.} We decrease the attenuation and the optical power of pulses in the Sagnac loop is about -20 dBm. We use a power meter to measure the optical power at both ports of the PBS. We measured the optical power once every 100 ms and tested it for 1000 seconds.} As shown in Fig.~S3, the blue points mean the ratio of slow-axis to fast-axis and the minimum value is still over 26.5 dB. As we only need 100 seconds to perform a round of measurement, the ratio can be easily maintained above 26.5 dB.

\begin{figure}[h!]
	\includegraphics[width=\columnwidth]{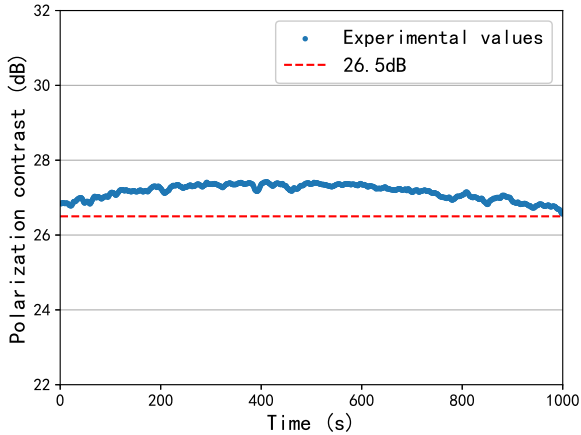}
	\caption{{The extinction ratio of pulse polarization versus time.} We measured the extinction ratio of polarization of Alice's and Bob's pulses by a polarization beam splitter (PBS) for 1000 seconds while our system considering realistic sources, only runs for 100 seconds. Its value is always above 26.5 dB. }
	\label{fig06}
\end{figure}

\subsection{Pattern effect}
Practical electro-optic modulators and electrical
drivers are band limited, which results in state
correlation between the optical pulses as well as intensity
fluctuation of individual pulses. {Pulse} correlation is particularly
critical for the security since current security analysis usually
assumes independent and identically distributed pulses. In this experiment, we do not need to modulate different intensities and we only need to quantify the phase correlation. {The experimental structure is same as that used for measuring phase shift. Here we only consider the influences of pattern effects limited to the adjacent pulses~\cite{Yoshino2018}.} Consequently, for four phases, sixteen patterns exist. We quantified the phase correlation by measuring the intensities under different patterns. Bob's PM is set at an unmodulated phase 0 and Alice modulates each pattern with equal probability. The modulated pulses and unmodulated pulses interfere at BS and the results are measured by D1 and D2 with detection efficiencies of 66.3\% and 73.6\%. Denote phase 0 as $S_1$, phase $\pi/2$ as $S_2$, phase $\pi$ as $S_3$ and phase $3\pi/2$ as $S_4$. The pattern effects of phase 0 and $\pi/2$ are given by D1's detection results while the pattern effects of phase $\pi$ and $3\pi/2$ are given by D2's detection results. As shown in Table~S2, the maximum value, or say sin$\psi$, is $8.54\times 10^{-3}$. Note that here we only measured the deviation from average values while theoretically we should measure the largest intensity deviation of individual pulses in each pattern. It is a remaining problem for future study.

\begin{table}[h!]
	\caption{{Detection clicks of pulses with different phases for sixteen types of predecessors.} }
	\setlength{\tabcolsep}{0.56cm}  
	\begin{tabular}
		{ccccc}  \hline \hline
		Pattern &  Counts of & deviation from \\ 
		& second pulse & average value \\ \hline
		$S_1 \rightarrow S_1$ & 411665 & $1.54\times 10^{-3}$ \\ 
		$S_2 \rightarrow S_1$ & 412610 & $7.52\times 10^{-4}$ \\ 
		$S_3 \rightarrow S_1$ & 411939 & $8.76\times 10^{-4}$ \\ 
		$S_4 \rightarrow S_1$ & 412986 & $1.66\times 10^{-3}$ \\ \hline
		$S_1 \rightarrow S_2$ & 206867 & $5.61\times 10^{-4}$ \\ 
		$S_2 \rightarrow S_2$ & 208191 & $6.96\times 10^{-3}$ \\ 
		$S_3 \rightarrow S_2$ & 206554 & $8.08\times 10^{-4}$ \\ 
		$S_4 \rightarrow S_2$ & 205362 & $6.72\times 10^{-3}$ \\ \hline
		$S_1 \rightarrow S_3$ & 455404 & $2.93\times 10^{-3}$ \\ 
		$S_2 \rightarrow S_3$ & 454520 & $9.83\times 10^{-4}$ \\ 
		$S_3 \rightarrow S_3$ & 452736 & $2.95\times 10^{-3}$ \\ 
		$S_4 \rightarrow S_3$ & 453635 & $9.66\times 10^{-4}$ \\ \hline
		$S_1 \rightarrow S_4$ & 227822 & $6.37\times 10^{-3}$ \\ 
		$S_2 \rightarrow S_4$ & 227191 & $3.58\times 10^{-3}$ \\ 
		$S_3 \rightarrow S_4$ & 226059 & $1.42\times 10^{-3}$ \\ 
		$S_4 \rightarrow S_4$ & 224446 & $8.54\times 10^{-3}$ \\ 
		\hline \hline
	\end{tabular}
	
	\label{tab06}
\end{table}

\section{Scheme with independent lasers}
We extend the plug-and-play system to a scheme with independent lasers in this part. As shown in Fig.~S4, this system includes two independent users and an untrusted intermediate node.\\

\begin{figure*}[t!]
	\includegraphics[width=2\columnwidth]{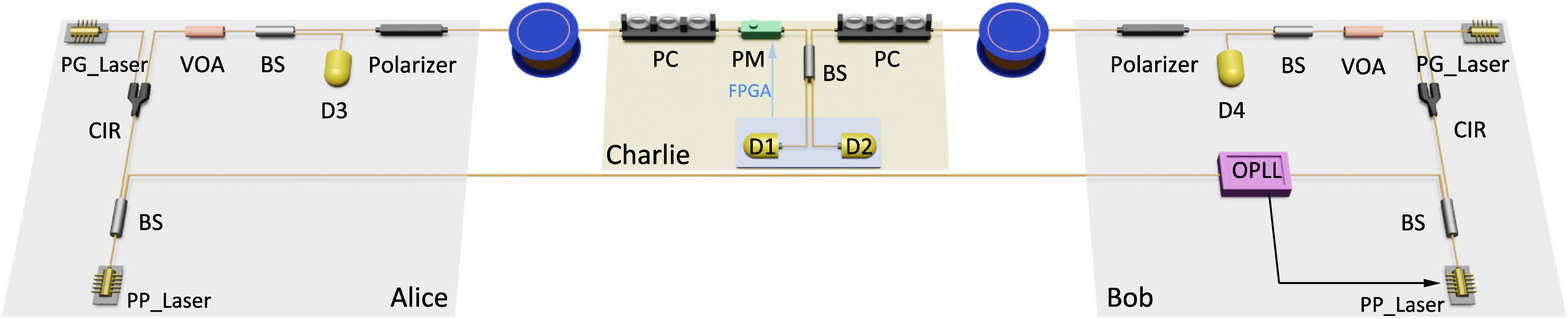}
	\caption{{Experimental setup of two independent users.} The phase-modulated pulses are generated by light sources without phase modulators. The light source consists of a phase-preparation laser (PP-Laser) and a pulse-generation laser (PG-Laser). PG-Laser is connected to PP-Laser via an optical circulator (CIR). Bob owns an optical phase-locked loop (OPLL) to lock his PP-Laser to Alice's PP-Laser. After traveling through quantum channels, Alice’s and Bob’s encoded pulses interfere with each other on Charlie’s beam splitters (BS), and are finally detected by two superconducting nanowire single-photon detectors, D1 and D2. EPC, electrical polarization controller; VOA, variable optical attenuator; PC, polarization controller; PM, phase modulator. }
	\label{fig07}
\end{figure*}

The phase-modulated pulses are directly generated employing the idea of optical injection~\cite{yuan2016directly,de2021real}, which has been proven to effectively remove the need for a phase modulator. Additionally, without multi-intensity modulation, our scheme does not need an intensity modulator. Consequently, Eve cannot get any information about phase modulation and intensity modulation from the backscattered light and thus this scheme can resist Trojan horse attacks. As shown in Fig.~S4, Alice's (Bob) laser source is made up with a phase-preparation laser (PP-Laser) and a pulse-generation laser (PG-Laser).
Two laser diodes are connected with an optical circulator. PP-Laser acts as a master laser and PG-Laser acts as a slave laser. {PP-Laser is always biased over its threshold to generate continuous-wave light. PG-Laser is gain-switched corresponding with modulation signals applied to PP-Laser for avoiding side effects and sends pulses inheriting the phase prepared by PP-Laser. By introducing perturbations in the driving signal of PP-Laser, the relative phase between two secondary pulses can be set to an arbitrary value. To illustrate this point more specifically, consider a continuous wave laser above threshold with the central frequency $\nu_0$. When applying a perturbation on the driving signal, the optical frequency shifts by $\Delta\nu$. The signal duration is $t_0$. After $t_0$, the frequency is restored to the initial value. The phase difference caused by this signal is $2\pi\Delta \nu t_0$.} Bob locks his PP-Laser to Alice’s PP-Laser by utilizing an optical phase-locked loop (OPLL). The modulated optical pulses are then attenuated to the single-photon level by a variable optical attenuator and their intensities are monitored by detectors. A polarizer is used to improve the polarization extinction ratio of pulses sent.\\
After traveling through optical fibers, Alice’s and Bob’s pulses interfere with each other on Charlie’s BS. Two polarization controllers (PCs) are used to modify the polarization of the incident pulses to achieve the maximum detection efficiencies. One feedback phase modulator (PM) is added in the path
from Alice to Charlie’s BS to compensate for the relative phase drift introduced by fiber channels.The interference results are finally detected by two superconducting single-photon 
detectors, D1 and D2.\\
Since our protocol does not need intensity modulation, this scheme removes the need for electro-optic modulators and electrical drivers, simplifying the experiment.  Here, we propose a scheme that two independent parties share secure keys considering source flaws with developed technology, which can be widely implemented in the approaching quantum networks.

\section{Experimental results}
\begin{table}[h!]
	\caption{Efficiencies of measurement station.}
	\setlength{\tabcolsep}{1.1cm}  
	\begin{tabular}
		{cc}  \hline \hline
		Optical devices &  Attenuation \\ \hline
		CIR 2$\rightarrow$3  & 0.62 dB \\ \hline
		BS-1  & 0.45 dB \\ \hline
		BS-2  & 0.47 dB \\ \hline
		PC-1  & 0.16 dB \\ \hline
		PC-2  & 0.18 dB \\ \hline

	\end{tabular}
	
	\label{tab07}
\end{table}
\begin{table}[h]
	\caption{Experimental results obtained under different losses.} \label{tab08}
	\setlength{\tabcolsep}{0.05cm}  
	\begin{tabular}{c|c|c|c|c|c}
		\hline\xrowht[()]{8pt}
		
		Channel loss & 10 dB & 16 dB & 20 dB & 30 dB & 35 dB\\
		\hline \xrowht[()]{8pt}
		N  & $10^{10}$ & $10^{10}$ & $10^{10}$ & $10^{10}$& $ 10^{10}$ \\
		\hline\xrowht[()]{8pt}
		$E_p^{*}$ & {18.4}\% &{19.4}\% & {19.0}\%& {29.4}\%& {35.5}\% \\
		\hline\xrowht[()]{8pt}
		$E_p$ & {23.7}\% & {30.8}\% & {39.3}\% &50.0\% &50.0\%\\
		\hline\xrowht[()]{8pt}
		Detected 00 (D1) & 11697938 & 2700313  & 882175 & 108166& 29340\\
		\hline\xrowht[()]{8pt}
		Detected 0$\pi$ (D1)& 36848 & 9031& 2703 & 261 & 113 \\
		\hline\xrowht[()]{8pt}
		Detected $\pi$0 (D1)& 23186 & 6897 & 1994 & 319 &73\\
		\hline\xrowht[()]{8pt}
		Detected $\pi\pi$ (D1)&11633036 & 2699790 & 882851 & 108459 &29303\\
		\hline\xrowht[()]{8pt}
		Detected $\frac{\pi}{2}\frac{\pi}{2}$ (D1)& 143957 & 33188 &10872 & 1773 &472\\
		\hline\xrowht[()]{8pt}
		Detected $\frac{\pi}{2}\frac{3\pi}{2}$ (D1)& 491 & 108  & 35& 5 &3\\
		\hline\xrowht[()]{8pt}
		Detected $\frac{3\pi}{2}\frac{\pi}{2}$ (D1)& 603 &162&  45 & 8 &3\\
		\hline\xrowht[()]{8pt}
		Detected $\frac{3\pi}{2}\frac{3\pi}{2}$ (D1)& 143048 & 33308 & 10766& 1284 & 331\\
		\hline\xrowht[()]{8pt}
		Detected 00 (D2)  & 7516 &1693  & 570 & 106 & 65\\
		\hline\xrowht[()]{8pt}
		Detected 0$\pi$ (D2) & 12982085 & 2998179 & 973757& 121242 &31961\\
		\hline\xrowht[()]{8pt}
		Detected $\pi$0 (D2)& 12973080 & 2996549 & 971129& 125848 & 33099\\
		\hline\xrowht[()]{8pt}
		Detected $\pi\pi$ (D2) & 45002 & 8655 & 2994 &590 &163\\
		\hline\xrowht[()]{8pt}
		Detected $\frac{\pi}{2}\frac{\pi}{2}$ (D2)& 347 & 60 & 20 &4 &1\\
		\hline\xrowht[()]{8pt}
		Detected $\frac{\pi}{2}\frac{3\pi}{2}$ (D2)& 167242 & 38832& 12464& 1543 & 408\\
		\hline\xrowht[()]{8pt}
		Detected $\frac{3\pi}{2}\frac{\pi}{2}$ (D2)& 167263 & 38658 & 12650& 1348& 346\\
		\hline\xrowht[()]{8pt}
		Detected $\frac{3\pi}{2}\frac{3\pi}{2}$ (D2)& 823 & 216 & 71 & 8& 2\\
		\hline		
	\end{tabular}
\end{table}We summarized the optical element transmittance in Table~S3. The optical elements include the polarization controllers (PCs), circulator (CIR), and the beam splitter (BS). The results are given for each output (D1/D2) as appropriate.\\
The experimental results are summarized in Table~S4, the
total pulses sent in the experiment N, the phase error rate regardless the source flaws $E_{\rm P}^{*}$, the phase error rate considering the source flaws $E_{\rm P}$, the number of detected pulses is labeled as “Detected-AB (Di)”, where “A” (“B”) means an A (B) phase was added on the pulses by Alice (Bob) and detected by Di, i $\in \{1, 2\}$. For 30 dB and 35 dB, the key rates are 0 and $E_{\rm P} = 50\%$.


\end{document}